\documentclass[reprint, double column, superscriptaddress,prl, showkeys]{revtex4-1}

\usepackage{float}
\usepackage{graphicx,epsfig}
\usepackage{amssymb}
\usepackage{amsmath}
\usepackage{bm}
\usepackage{braket}

\usepackage{makecell}
\usepackage{textcomp}
\usepackage{color}
\usepackage{pgffor}
\usepackage{pdfpages}

\usepackage{soul}

\usepackage[bookmarksnumbered,bookmarksopen]{hyperref}
\hypersetup{
   colorlinks=true,       
}

\usepackage{BOONDOX-cal}

\makeatletter

\AtBeginDocument{\let\LS@rot\@undefined}
\makeatother
\newif\ifarXiv
\arXivtrue

\usepackage{soul}

\begin{document}
\setcounter{page}{1}

\title[]{Observation of a Stripe/nematic Phase of Composite Fermions}
\author{Chengyu \surname{Wang}}
\author{S. K. \surname{Singh}}
\author{C. T. \surname{Tai}}
\author{A. \surname{Gupta}}
\author{L. N. \surname{Pfeiffer}}
\author{K. W. \surname{Baldwin}}
\author{M. \surname{Shayegan}}
\affiliation{Department of Electrical and Computer Engineering, Princeton University, Princeton, New Jersey 08544, USA}

\date{\today}

\begin{abstract}
{Electronic stripe/nematic phases are fascinating strongly-correlated states characterized by spontaneous rotational symmetry breaking. In the quantum Hall regime, such phases typically emerge at half-filled, high-orbital-index ($N\geq2$) Landau levels (LLs) where the short-range Coulomb interaction is softened by the nodes of electron wave functions. In the lowest ($N=0$) LLs, these phases are not expected. Instead, composite fermion (CF) liquids and fractional quantum Hall states, which are well explained in the picture of weakly interacting CF quasiparticles, are favored. Here we report the observation of an unexpected stripe/nematic phase in the \textit{lowest} LL at filling factor $\nu=5/8$ in ultrahigh-quality GaAs two-dimensional \textit{hole} systems, evinced by a pronounced in-plane transport anisotropy. Remarkably, $\nu=5/8$ can be mapped to a half-filled, high-index CF LL ($N_{\text{CF}}=2$), analogous to the $N=2$ hole LL. Our finding signals a novel stripe/nematic phase of CFs, driven by the residual long-range interaction among these emergent quasiparticles. This phase is surprisingly robust, surviving up to $\sim$100 mK. Its absence in electron-type systems suggests that severe LL mixing stemming from the large hole effective mass and non-linear LL fan diagram plays a crucial role in modifying the CF-CF interaction.}
\end{abstract}

\maketitle

The concept of quasiparticles is pivotal for understanding the complex electronic properties in condensed matter physics. In solids, the behavior of electrons in a periodic potential created by the atomic lattice can be effectively described using Bloch electrons -- quasiparticles that resemble free electrons but with a renormalized effective mass. In the extreme quantum limit of a two-dimensional electron system (2DES), where electrons partially occupy the lowest ($N=0$), highly-degenerate Landau level (LL), strong electron-electron Coulomb interaction leads to the emergence of exotic correlated phases, such as fractional quantum Hall states (FQHSs) \cite{Tsui.PRL.1982, Laughlin.PRL.1983, Halperin.Book.2020}. These many-body electronic phases can be phenomenologically understood through an effective single-particle framework by introducing a new type of quasiparticles known as composite fermions (CFs), which are formed by attaching an even number of magnetic flux quanta to each electron \cite{Jain.PRL.1989, Halperin.PRB.1993, Jain.Book.2007}.  The flux attachment reduces the external magnetic field and, in an effective mean-field theory, also screens the Coulomb interaction. As a result, two-flux CFs ($^2$CFs) experience zero effective magnetic field $B^*$ at LL filling factor $\nu=1/2$, and form a compressible Fermi sea. This has been confirmed experimentally \cite{Willett.PRL.1993, Kang.PRL.1993, Kamburov.PRL.2014}. As the system moves away from $\nu=1/2$, the $^2$CFs experience a finite $B^*=B-B_{1/2}$ and form their own LLs, known as Lambda levels ($\Lambda$Ls) \cite{Jain.Book.2007}. Similar to electrons, $^2$CFs exhibit integer QHSs when $\Lambda$Ls are fully occupied, which manifest as Jain-sequence FQHSs at $\nu=n/(2n\pm1)$, $n=1,2,3...$. 

Electron-electron interaction in excited ($N\geq1$) LLs, however, is rather different, because the nodes of electron wavefunctions soften the short-range part of the Coulomb interaction, giving rise to exotic correlated states which are not favored in the $N=0$ LLs. One peculiar example is the candidate non-Abelian FQHS observed in half filled $N=1$ LL at $\nu=5/2$ \cite{Willett.PRL.1987}, which is believed to originate from BCS-like CF pairing \cite{Moore.NPB.1991, Read.PRB.2000, Sharma.PRB.2020}. In higher $N\geq2$ LLs, stripe or electronic versions of liquid-crystal-like phases, characterized by significant in-plane transport anisotropy emerge at half fillings, e.g. at $\nu=9/2$ \cite{Koulakov.PRL.1996, Fogler.PRB.1996, Moessner.PRB.1996, Lilly.PRL.1999, Du.SSC.1999, Fradkin.PRB.1999, Fradkin.PRL.2000, Oganesyan.PRB.2001, Fradkin.review.2010, Lee.PRL.2018}. Early Hartree-Fock theories predicted that these phases stem from unidirectional charge-density waves consisting of stripes with alternating integer $\nu$ (e.g, $\nu=4$ and 5) \cite{Koulakov.PRL.1996, Fogler.PRB.1996, Moessner.PRB.1996, Sammon.PRB.2019}. At finite temperatures, and in the presence of quantum fluctuations and disorder, the stripe order can be disrupted, leading to nematic phases \cite{Fradkin.PRB.1999, Fradkin.PRL.2000, Fradkin.review.2010}. In the remainder of the paper, we refer to such phases as stripe/nematic (S/N) phases. In a more general picture, nematic orders can also arise from Pomeranchuk instability of Fermi seas \cite{Oganesyan.PRB.2001, Lee.PRL.2018}.

\begin{figure*}[t]
  \begin{center}
    \psfig{file=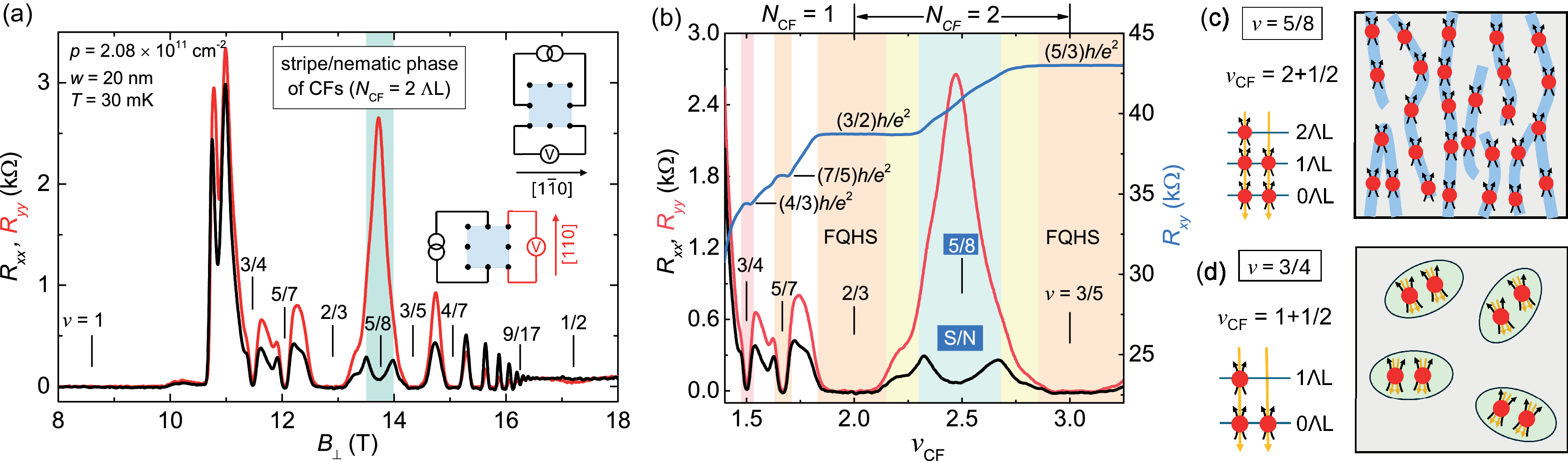, width=1\textwidth}
  \end{center}
  \caption{\label{fullfield} 
   {\bf An S/N phase of CFs in the lowest LL.} (a) Longitudinal resistances $R_{xx}$ and $R_{yy}$ vs perpendicular magnetic field $B_{\perp}$ measured along two mutually perpendicular crystal directions at $T\simeq30$ mK. The circuit configurations used for the measurements are shown in the right insets. Our 2DHS exhibits a highly-anisotropic behavior between two Jain-sequence FQHSs $\nu=2/3$ and 3/5, showing a peak anisotropy ($R_{yy}/R_{xx}\simeq 40$) near $\nu=5/8$. This signals the emergence of an S/N phase in the FQHS regime. (b) $R_{xx}$, $R_{yy}$, and Hall resistance $R_{xy}$ are plotted as a function of CF filling factor, $\nu_{\text{CF}}$. FQHSs identified by vanishing $R_{xx}$ and $R_{yy}$, and accompanied by quantized Hall plateaus, are observed at $\nu=3/5$, 2/3, 5/7, and 3/4, corresponding to $\nu_{\text{CF}}=3$, 2, 5/3, and 3/2, respectively. No $R_{xy}$ plateau is seen near $\nu=5/8$ ($\nu_{\text{CF}}=5/2$). (c) Origin of the S/N phase at $\nu=5/8$: First, we map hole LL filling $\nu=5/8$ to CF $\Lambda$L filling $\nu_{\text{CF}}=5/2$, where CFs fully occupy the $N_{\text{CF}}=0$ and $N_{\text{CF}}=1$ $\Lambda$Ls, and half occupy the topmost, $N_{\text{CF}}=2$ $\Lambda$L. A S/N phase of CFs forms in the half-filled $N_{\text{CF}}=2$ $\Lambda$L, analogous to electronic S/N phases in high LLs, e.g. at $\nu=$ 9/2. (d) Origin of the exotic $\nu=$ 3/4 FQHS: Paired FQHS of CFs in the half-filled $N_{\text{CF}}=1$ $\Lambda$L. We assume CFs are fully spin-polarized; this is reasonable, given the very large $B_{\perp}$ where our observations are made.
   }
  \label{fig:fullfield}
\end{figure*}

We report here the observation of an unusual, \textit{anisotropic} phase at $\nu=$ 5/8 in the \textit{lowest} LL of ultrahigh-quality GaAs 2D hole systems (2DHSs). The LL filling $\nu=$ 5/8 can be mapped to a \textit{hole-flux} CF $\Lambda$L filling $\nu_{\text{CF}}=$ 5/2, representing a half-occupied $N_{\text{CF}}=2$ $\Lambda$L on top of two, fully occupied, lower $\Lambda$Ls ($N_{\text{CF}}=0$ and 1). The exotic phase we observe at $\nu=5/8$ is therefore very likely a manifestation of an S/N phase of \textit{interacting CFs}; see Fig. \ref{fig:fullfield}(c). This phase is novel and intricate because both the S/N phase and the CF quasiparticles forming this phase have collective origins.

We studied ultrahigh-quality 2DHSs confined to GaAs quantum wells grown on GaAs (001) substrates by molecular beam epitaxy. They were grown following the optimization of the growth chamber vacuum integrity and the purity of the source materials \cite{Chung.NatMater.2021}, as well as an optimized, stepped-barrier design \cite{Chung.PRM.2022, Gupta.PRM.2024}. We performed our experiments on $4\times 4$ mm$^2$ van der Pauw geometry samples cleaved from a 2-inch GaAs wafer, with alloyed In:Zn contacts at the four corners and side midpoints. The samples were cooled in a dilution refrigerator with a base temperature of $\simeq 30$ mK. We measured the longitudinal resistances along [1$\bar{1}$0] ($R_{xx}$) and [110] ($R_{yy}$) crystal directions \cite{Footnote.direction}, and the Hall resistance $R_{xy}$ using the conventional, low-frequency ($\sim17$ Hz), lock-in amplifier technique.

We focus on transport measurements on a GaAs 2DHS in the lowest LL ($\nu<1$). Figure \ref{fig:fullfield}(a) presents $R_{xx}$ and $R_{yy}$ vs perpendicular magnetic field $B_{\perp}$. We observe a dramatic anisotropic behavior at $\nu=5/8$: $R_{xx}$ exhibits a deep minimum while $R_{yy}$ shows a maximum (with $R_{yy}/R_{xx}\simeq40$), reminiscent of the S/N phases that typically emerge in high ($N\geq2$) LLs \cite{Lilly.PRL.1999, Du.SSC.1999, Fradkin.review.2010}. Such anisotropic behavior at $\nu=5/8$ is also seen in two other samples with similar 2D hole densities, but different quantum well widths; see Supplemental Material (SM) Fig. S5 \cite{SM}. As we elaborate later, this anisotropic behavior signals the emergence of an S/N phase of interacting $^2$CFs in a half-filled, high ($N_{\text{CF}}=2$) $\Lambda$L. The S/N phase at $\nu=5/8$ is in stark contrast to the nearly isotropic behavior we observe elsewhere in the lowest LL ($\nu<1$). Near $\nu=$ 1/2, $R_{xx}$ and $R_{yy}$ are featureless and have similar values, consistent with an isotropic Fermi sea of $^2$CFs. On the lower-$B_{\perp}$ side of $\nu=1/2$, a sequence of minima which signal the Jain-sequence FQHSs are observed at $\nu=2/3$, 3/5, 4/7, ..., up to 9/17. It is worth noting that, between $\nu=1$ and 2/3, an \textit{even-denominator} FQHS emerges at $\nu=3/4$, consistent with what was recently reported in ultrahigh-quality GaAs 2DHSs \cite{Wang.PRL.2022, Wang.PNAS.2023}. These observations collectively demonstrate the exceptionally high quality of the GaAs 2DHS in our study.

\begin{figure*}[t]
  \begin{center}
    \psfig{file=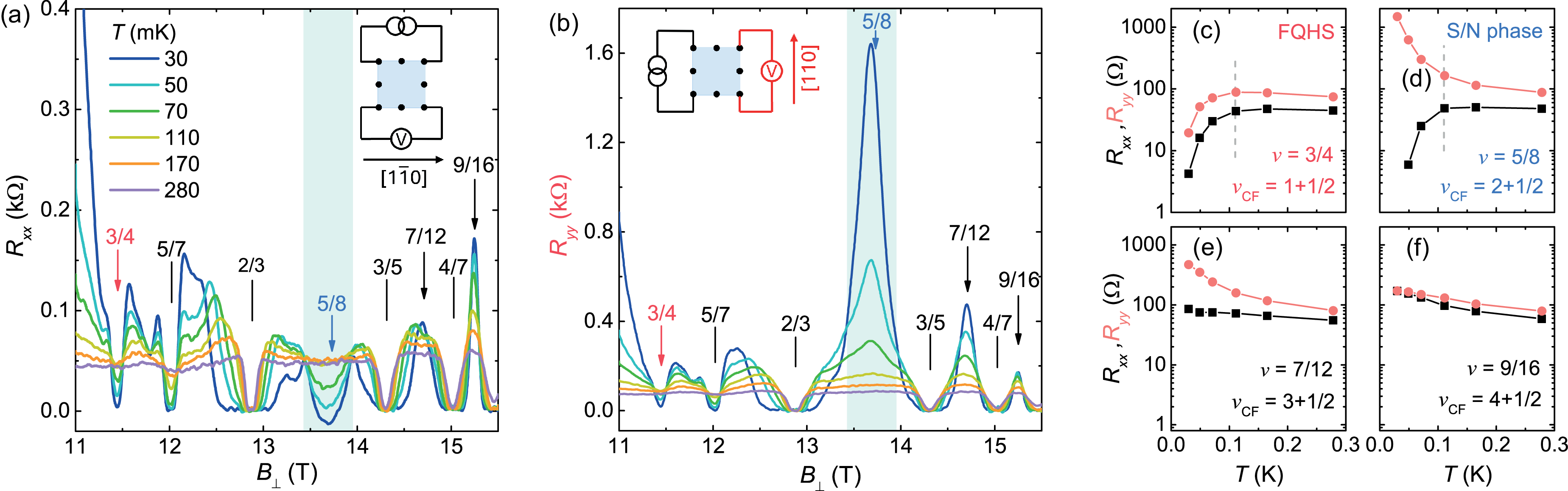, width=1 \textwidth}
  \end{center}
  \caption{\label{Tdep}
    {\bf Temperature dependence.} (a, b) $R_{xx}$ and $R_{yy}$ vs $B_{\perp}$ traces measured at different $T$. Vertical black lines mark the $B_{\perp}$ positions of odd-denominator FQHSs. Filling factors $\nu=3/4$, 5/8, 7/12, and 9/16, which correspond to half-integer $\nu_{\text{CF}}=$ 3/2, 5/2, 7/2, and 9/2, are marked with arrows. Insets show the circuit configurations used for the measurements. (c-f) $R_{xx}$ and $R_{yy}$ vs $T$ at $\nu=$ 3/4, 5/8, 7/12, and 9/16. At $\nu=3/4$, as we lower $T$, both $R_{xx}$ and $R_{yy}$  first slightly increase, and then decrease at $T\lesssim 0.1$ K when a FQHS is developing. At $\nu=7/12$ and 9/16, both $R_{xx}$ and $R_{yy}$ monotonically increase with decreasing $T$. In contrast to these fillings where $R_{xx}$ and $R_{yy}$ exhibit the same trend with $T$, at $\nu=5/8$, $R_{xx}$ and $R_{yy}$ deviate from each other and become highly anisotropic at $T\lesssim 0.1$ K, signaling the emergence of an S/N phase.
    }
  \label{fig:Tdep}
\end{figure*}

To shed light on the origin of the exotic S/N phase at $\nu=5/8$, we present our data in the $^2$CF picture by plotting $R_{xx}$, $R_{yy}$, and $R_{xy}$ in Fig. \ref{fig:fullfield}(b) as a function of $\nu_{\text{CF}}$, where $\nu_{\text{CF}}$ is the CF $\Lambda$L filling factor obtained from the relation $\nu=\nu_{\text{CF}}/(2\nu_{\text{CF}}+1)$. The Jain-sequence FQHSs at $\nu=2/3$ and 3/5, evinced by wide, quantized $R_{xy}$ plateaus accompanied with vanishing $R_{xx}$ and $R_{yy}$, can be interpreted as integer QHSs of $^2$CFs with $\nu_{\text{CF}}=2$ and 3. An S/N phase is observed between $\nu=2/3$ and 3/5 (between $\nu_{\text{CF}}=2$ and 3). Here CFs fully occupy the $N_{\text{CF}}=0$ and $N_{\text{CF}}=1$ $\Lambda$Ls, and partially occupy the $N_{\text{CF}}=2$ $\Lambda$L, assuming CFs are fully spin polarized. The peak anisotropy is seen near $\nu=5/8$, corresponding to $\nu_{\text{CF}}=2+1/2$, where the topmost $N_{\text{CF}}=2$ $\Lambda$L is half occupied [Fig. \ref{fig:fullfield}(c)]. No $R_{xy}$ plateau is seen near $\nu=5/8$. These observations highly resemble the conventional S/N phases reported in high LLs of GaAs 2DESs \cite{Lilly.PRL.1999, Du.SSC.1999, Fradkin.review.2010}, suggesting that what we observe at $\nu=5/8$ is an S/N phase. Remarkably, this S/N phase is observed in the \textit{lowest} LL, emerging from interacting $^2$CFs rather than interacting holes.

Several other observations support our claim that CFs are interacting in a partially-filled $\Lambda$L in our 2DHS. The filling factor $\nu=3/4$, at which an exotic, \textit{even-denominator} FQHS is observed [see Fig. 1(b)], can be mapped to $\nu_{\text{CF}}=1+1/2$ [Fig. \ref{fig:fullfield}(d)]. A likely origin of this FQHS is the CF-CF interaction and pairing in the half-filled $N_{\text{CF}}=1$ $\Lambda$L \cite{Wang.PRL.2022, Wang.PNAS.2023}. Such CF-CF interaction resembles the electron-electron interaction in the $N=1$ LLs, which is believed to be the key to stabilize paired, non-Abelian FQHSs at even-denominator fillings, e.g. at $\nu=5/2$ \cite{Willett.PRL.1987}. We also note that there are two inflection points in both $R_{xx}$ and $R_{yy}$ between $\nu=2/3$ and 3/5, where $R_{xy}$ shows quantization merging into the $\nu=2/3$ and 3/5 plateaus [see the yellow-shaded regions in Fig. \ref{fig:fullfield}(b)]. Similar features are also seen near $\nu=2/3$ in other two samples; see SM Figs. S1 and S2 for details \cite{SM}. These features are possibly indications of developing \textit{reentrant} FQHSs, which are believed to be bubble phases of interacting CFs \cite{Lee.PRB.2002, Goerbig.PRL.2004, Shingla.NatPhys.2023}. The $\nu_{\text{CF}}$ positions of these features, $\simeq 2+0.25$ and $2+0.75$, are close to the $\nu$ positions of bubble phases observed in the $N=2$ LL \cite{Ro.PRB.2020}, suggesting that CF-CF interaction in the $N_{\text{CF}}=2$ $\Lambda$L is analogous to the electron-electron interaction in the $N=2$ LL.

Figure 2 displays the temperature dependence data, providing further evidence for the emergence of an S/N phase at $\nu=5/8$; see SM Fig. S6 for the temperature dependence of two other samples \cite{SM}. In Figs. \ref{fig:Tdep}(a, b), we present $R_{xx}$ and $R_{yy}$ vs $B_{\perp}$ traces \cite{Footnote.config, Simon.PRL.1999} measured at different $T$; see SM Fig. S1 for $R_{xy}$ data \cite{SM}. As we increase $T$ from 30 mK, the $R_{xx}$ minimum and $R_{yy}$ maximum at $\nu=$ 5/8 weaken sharply. Both features disappear at $T\simeq$ 110 mK, and exhibit little temperature dependence as we further increase $T$ to 280 mK. These trends are more clearly revealed in Fig. \ref{fig:Tdep}(d). Our observations signal the thermal melting of the S/N phase at $\nu=5/8$, strikingly resembling what was reported for S/N phases in high LLs of GaAs 2DESs \cite{Lilly.PRL.1999, Du.SSC.1999, Cooper.PRB.2002, Fradkin.review.2010}.

In Figs. \ref{fig:Tdep}(c,e,f), we show the temperature dependence of $R_{xx}$ and $R_{yy}$ at other fillings $\nu=$ 3/4, 7/12, and 9/16. Similar to $\nu=5/8$ which corresponds to $\nu_{\text{CF}}=$ 2+1/2, these $\nu$ can also be mapped to half-integer $\nu_{\text{CF}}$. In contrast to the anisotropic behavior at $\nu=5/8$, $R_{xx}$ and $R_{yy}$ qualitatively follow the same trend. At $\nu=3/4$ ($\nu_{\text{CF}}=1+1/2$), where the ground state is a FQHS, both $R_{xx}$ and $R_{yy}$ approach zero at low $T$ [Fig. \ref{fig:Tdep}(c)] \cite{Wang.PRL.2022}. At $\nu=7/12$ and 9/16 ($\nu_{\text{CF}}=3+1/2$ and $4+1/2$), both $R_{xx}$ and $R_{yy}$ increase slightly at lower $T$ [Figs. \ref{fig:Tdep}(e, f)] \cite{Footnote.7/12}, consistent with what was reported at FQH plateau-to-plateau transitions \cite{Madathil.PRL.2023}. 

\begin{figure*}[t]
  \begin{center}
    \psfig{file=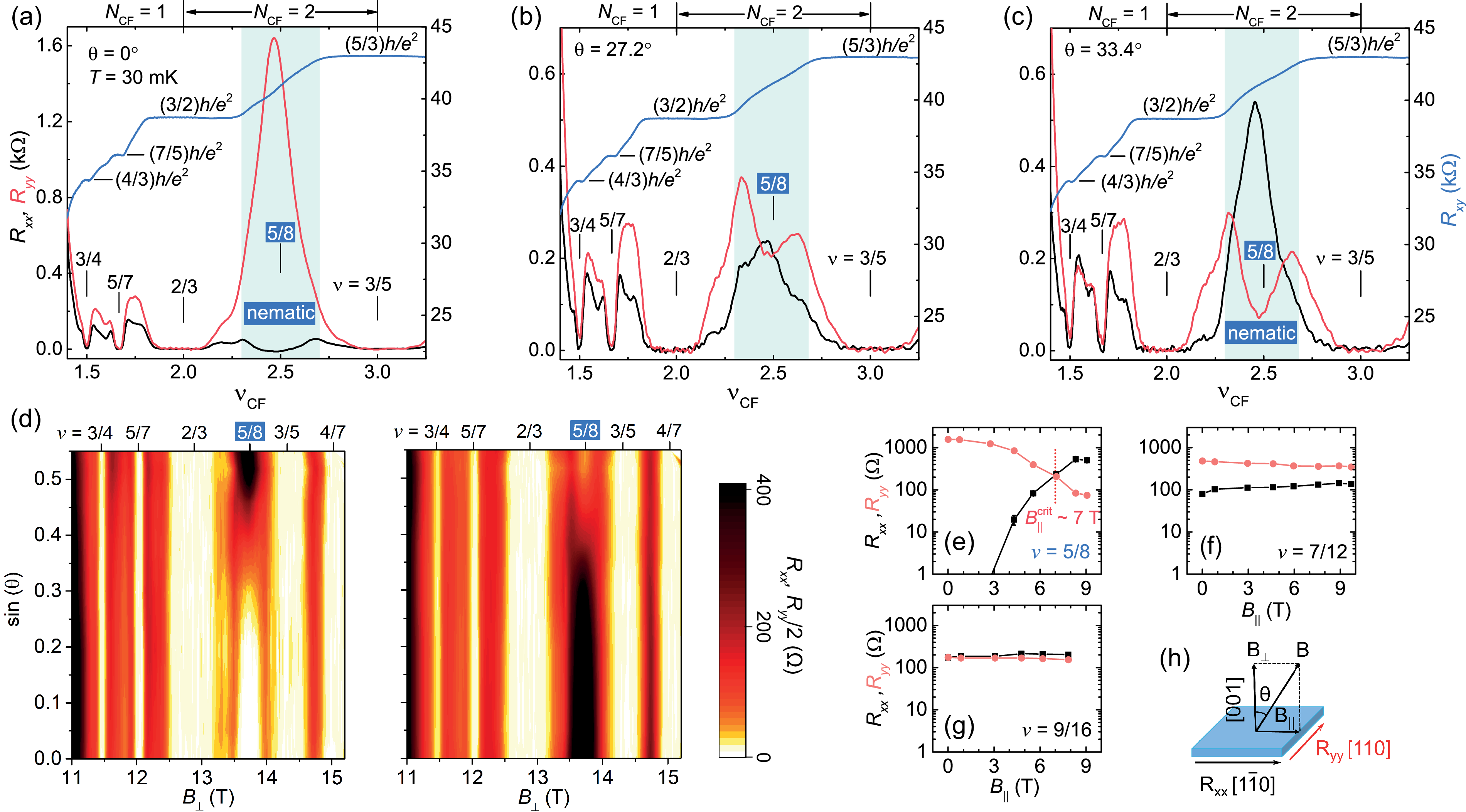, width=1 \textwidth}
  \end{center}
  \caption{\label{tilt}
    {\bf Role of in-plane magnetic field.} (a-c) $R_{xx}$, $R_{yy}$, and $R_{xy}$ vs $\nu_{\text{CF}}$ traces measured at (a) $\theta=0^{\circ}$, (b) $\theta=27.2^{\circ}$, and (c) $\theta=33.4^{\circ}$. (d) Color-scale plot of $R_{xx}$ (left) and $R_{yy}/2$ (right) as a function of $B_{\perp}$ and sin($\theta$). (e-g) $R_{xx}$ and $R_{yy}$ vs $B_{\parallel}$ at $\nu=$ 5/8, 7/12, and 9/16. The circuit configurations for $R_{xx}$ and $R_{yy}$ measurements are the same as shown in Fig. \ref{fig:Tdep}. (h) A schematic of the experimental setup.
    }
  \label{fig:tilt}
\end{figure*}

Next we study the evolution of the CF S/N phase in tilted $B$. The sample is mounted on a rotating stage to support \textit{in-situ} tilt; $\theta$ is the angle between $B$ and $B_{\perp}$; see Fig. \ref{fig:tilt}(h). The in-plane magnetic field $B_{\parallel}=B$sin($\theta$) is applied along the [1$\bar{1}$0] ($R_{xx}$) direction; see SM for data with $B_{\parallel}$ applied along the [110] ($R_{yy}$) direction \cite{SM}. Figures \ref{fig:tilt}(a-c) display data measured at three different $\theta$. At $\theta=0$, we observe an S/N phase at $\nu=$ 5/8 with $R_{yy}>>R_{xx}$. At $\theta=27.2^{\circ}$, $R_{xx}$ and $R_{yy}$ values at $\nu=$ 5/8 become very close. At the largest $\theta=33.4^{\circ}$, an S/N phase emerges again, but now $R_{xx}>>R_{yy}$. This evolution with $\theta$ is summarized in Fig. \ref{fig:tilt}(d), where we present color-scale plots of $R_{xx}$ and $R_{yy}$ as a function of $B_{\perp}$ and sin($\theta$). With increasing sin($\theta$), the S/N phase at $\nu=$ 5/8 gradually weakens, disappears at intermediate $\theta$, and eventually reappears at large $\theta$ with the hard axis switched to the $R_{xx}$ direction. Qualitatively similar behavior was also reported in high LLs of GaAs 2DESs \cite{Lilly.PRL.1999.tilt, Cooper.PRB.2002}, where S/N phases generally align their hard axis to the direction of $B_{\parallel}$ when $B_{\parallel}>B_{\parallel}^{\text{crit}}$. However, there are two surprises: (i) We observe a $B_{\parallel}^{\text{crit}}\simeq7$ T [Fig. \ref{fig:tilt}(e)], significantly larger than the typical values ($B_{\parallel}^{\text{crit}}\lesssim1$ T) reported for high LL S/N phases in GaAs 2DESs \cite{Lilly.PRL.1999.tilt, Cooper.PRB.2002}. (ii) At high $\theta$, the $\nu=5/8$ state is more fragile than what we observe at $\theta=0$: the anisotropic behavior disappears at a lower $T\simeq70$ mK (SM Fig. S2 \cite{SM}). This is in contrast to the S/N phases in high LLs which survive up to higher temperatures in the presence of a large $B_{\parallel}$ \cite{Cooper.PRB.2002}.

Unlike the S/N phase at $\nu=$ 5/8, other states in the lowest LL are rather insensitive to $\theta$. The FQHSs observed at $\nu$= 3/4, 5/7, 2/3, 3/5, and 4/7 remain robust with increasing $\theta$ [Figs. \ref{fig:tilt}(a-d)], suggesting that these FQHSs are fully spin polarized. This corroborates our assumption that CFs are fully polarized. At $\nu=$ 7/12 and 9/16 (corresponding to $\nu_{\text{CF}}=$ 7/2 and 9/2, respectively), $R_{xx}$ and $R_{yy}$ largely remain constant as a function of $B_{\parallel}$ [Figs. \ref{fig:tilt}(f,g)], indicating the absence of S/N instability at these fillings.

Our observation of a CF S/N phase highlights the important role of residual CF-CF interaction in the FQH regime. Similar to electrons in a high-orbital-index LL, CFs in a high-index $\Lambda$L also have nodes in their wavefunction \cite{Jain.PRL.1989, Jain.Book.2007}. Intuitively, the nodes in the CF wavefunction can soften the short-range residual Coulomb repulsion which is already very weak after forming CFs via flux attachment. Calculations of pseudopotentials for CF-CF interaction indeed suggest that residual CF-CF interaction in excited ($N_{\text{CF}}\geq1$) $\Lambda$Ls is often attractive \cite{Lee.PRB.2002, Lee.PRL.2001, Scarola.PRL.2002}. These calculations also predict that such residual CF-CF interaction can lead to numerous exotic correlated states including FQHSs at unconventional fillings, and stripe and bubble phases of CFs \cite{Lee.PRL.2001, Lee.PRB.2002, Scarola.PRL.2002}. FQHSs beyond the Jain sequence emerging from CF-CF interaction were reported in the lowest LL at both odd- and even-denominator fillings \cite{Goldman.SurfSci.1990, Pan.PRL.2003, Liu.PRL.2014, Kumar.NC.2018, Wang.PRL.2022, Wang.PNAS.2023}, e.g., at $\nu=4/11$ in GaAs 2DESs \cite{Pan.PRL.2003}, and at $\nu=3/4$ in GaAs 2DHSs \cite{Wang.PRL.2022, Wang.PNAS.2023}. An exotic bubble phase of CFs was also observed recently near $\nu=5/3$ in a GaAs 2DES \cite{Shingla.NatPhys.2023}. Note that these states emerge in the $N_{\text{CF}}=1$ CF $\Lambda$L, qualitatively similar to the many-body states observed in the $N=1$ electron LL. The S/N phase we observe at $\nu=5/8$, on the other hand, signals yet another exotic collective state emerging from CF-CF interaction in a higher ($N_{\text{CF}}=2$) CF $\Lambda$L.

While the potential for stripe formation in the FQH regime, as discussed in Refs. \cite{Lee.PRL.2001, Lee.PRB.2002}, offers a plausible explanation for our data, the robustness of the CF S/N phase we observe is intriguing. Collective states of CFs are generally considered exceedingly fragile because the CF-CF interaction is about an order of magnitude weaker than the electron-electron interaction (when LL mixing is ignored) \cite{Lee.PRL.2001, Lee.PRB.2002}. Now, the S/N phases in high LLs of GaAs 2DESs and 2DHSs are generally reported only at very low temperatures ($T\leq50$ mK) \cite{Lilly.PRL.1999, Du.SSC.1999, Manfra.PRL.2007}, suggesting even more stringent conditions for CF stripe formation. In contrast, the S/N phase we observe at $\nu=5/8$ persists up to $T\simeq$ 100 mK, indicating its surprising robustness. Such robustness, together with the $\nu=3/4$ FQHS we observe in the same sample, demonstrate that CFs in our 2DHS are highly interacting.

We attribute this surprisingly strong CF-CF interaction to the severe LL mixing (LLM) in our 2DHS samples. This seems counterintuitive at first sight since LLM softens the short-range component of the hole-hole Coulomb repulsion. However, the flux attachment required to form CFs is quantized and can lead to an \textit{overscreening}, rendering CFs in the $N_{\text{CF}}\geq1$ LLs mutually attractive \cite{Lee.PRB.2002, Lee.PRL.2001, Scarola.PRL.2002}. Therefore, the softening of hole-hole Coulomb repulsion by LLM actually results in an enhancement of residual CF-CF \textit{attractive} interaction, and can lead to a pairing of CFs and the formation of even-denominator FQHSs, e.g., at $\nu=3/4$ \cite{Wang.PRL.2022, Wang.PNAS.2023, Wang.PRL.2023, Zhao.PRL.2023}. Our calculated LL fan diagram (SM Fig. S4 \cite{SM}) shows that at $\nu=5/8$ the Coulomb energy is significantly larger than the LL separations, underscoring the importance of LLM. It is also worth noting that the S/N phase of CFs we observe at $\nu=5/8$, as well as the $\nu=3/4$ FQHS, are absent in ultrahigh-quality GaAs 2DESs, even though the mobility of GaAs 2DESs is an order of magnitude higher \cite{Chung.NatMater.2021}. Compared to 2DHSs, GaAs 2DESs have a much smaller effective mass and therefore experience negligible LLM at high $B_{\perp}$. This is consistent with our conjecture that LLM plays a crucial role in modifying the residual CF-CF interaction which stabilizes the exotic S/N phase of CFs.

\section*{Acknowledgement} 
We acknowledge support by the National Science Foundation (NSF) Grant No. DMR 2104771 for measurements, and the Gordon and Betty Moore Foundation’s EPiQS Initiative (Grant No. GBMF9615 to L.N.P.) for sample fabrication. Our measurements were partly performed at the National High Magnetic Field Laboratory (NHMFL), which is supported by the NSF Cooperative Agreement No. DMR 2128556, by the State of Florida, and by the DOE. This research was funded in part by QuantEmX grant from Institute for Complex Adaptive Matter and the Gordon and Betty Moore Foundation through Grant GBMF9616 to C. W., S. K. S., and C. T. T. We thank A. Bangura, G. Jones, R. Nowell and T. Murphy at NHMFL for technical assistance. We also thank Roland Winkler for the Landau level calculations, and Jainendra K. Jain for illuminating discussions.

\section*{Data availability} 
The data that support the findings of this article are openly available \cite{data}.



\section*{Supplementary material}

## I. SAMPLE PARAMETERS

Table I summarizes the sample parameters for the ultrahigh-quality GaAs two-dimensional hole systems (2DHSs) we used in this work. We studied six samples from five different wafers. All the data we present in the main text are measured for sample A1. Samples A1 and A2 are from the same wafer. Samples A (A1 and A2), B, and C have a similar 2D hole density of $p = 2.1$ (in units of $\times 10^{11}$ cm$^{-2}$, which we use throughout of the paper), but different quantum well (QW) widths (20 nm, 17.5 nm, and 15.75 nm). Samples D and E have the same QW widths (20 nm), but different 2D hole densities (1.3 and 0.6). The 2DHSs are confined to relatively narrow GaAs QWs, and only have the ground-state electric subband occupied. All the samples we study have transport mobility higher than $2\times 10^6$ cm$^2$/Vs. In this Supplemental Material, we provide additional data for sample A1, as well as data for other samples.

## II. TEMPERATURE DEPENDENCE OF HALL RESISTANCE

Figure S1 shows the temperature dependence of Hall traces for sample A1. We observe quantized Hall plateaus at Landau level (LL) filling factors $\nu = $ 3/4, 5/7, 2/3, 4/7, and 5/9, consistent with our notion of fractional quantum Hall states (FQHSs) at these fillings. Between $\nu = $ 2/3 and 3/5, we highlight the $B_\perp$ range where the anisotropic behavior is

TABLE I. **Sample parameters**

| Sample | density ($10^{11}$ cm$^{-2}$) | mobility ($10^6$ cm$^2$/Vs) | QW width (nm) | Al fraction for the barrier near the QW |
|---|---|---|---|---|
| A1 | 2.1 | 2.1 | 20 | 0.08 |
| A2 | 2.1 | 2.1 | 20 | 0.08 |
| B | 2.1 | 2.1 | 17.5 | 0.08 |
| C | 2.1 | 2.6 | 15.75 | 0.08 |
| D | 1.3 | 5.7 | 20 | 0.16 |
| E | 0.6 | 7.0 | 20 | 0.08 |

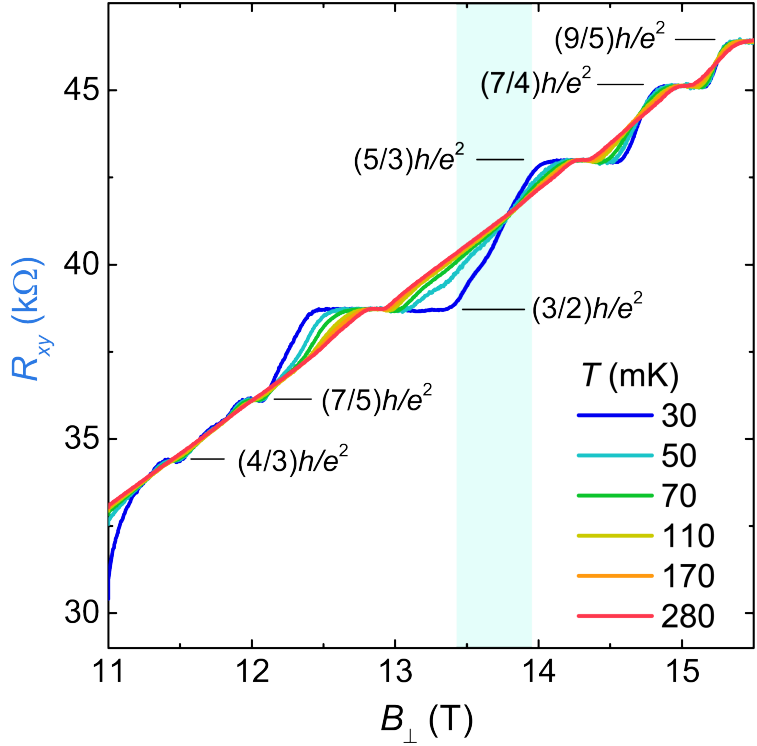

Fig. S1. **Temperature dependence of Hall traces.** $R_{xy}$ vs $B_\perp$ traces for sample A1 measured at $\theta = 0°$ at different temperatures.

observed in $R_{xx}$ and $R_{yy}$. Within this range, the $R_{xy}$ traces are smooth with no hints of plateau-like features.

## III. TEMPERATURE DEPENDENCE DATA IN TILTED MAGNETIC FIELD

As we show in Fig. 3 of the main text, in tilted $B$, the large in-plane magnetic field along $[1\bar{1}0]$ can switch the directions of hard- and easy-axes for the stripe/nematic (S/N) phase at $\nu = 5/8$. In Figs. S2(a,b), we present $R_{xx}$ and $R_{yy}$ traces measured at different temperatures in tilted field at $\theta = 35.5°$ (with $B_\parallel$ along $[1\bar{1}0]$ direction). At $T \simeq 30$ mK, we observe a peak in $R_{xx}$ and a minimum in $R_{yy}$ at $\nu = 5/8$, qualitatively similar to what we observe at $\theta = 33.4°$ [Fig. 3(c) in the main text]. Figure S2(c) shows the temperature dependence of $R_{xx}$ and $R_{yy}$ at $\nu = 5/8$ in tilted field at $\theta = 35.5°$. For $T < 70$ mK, $R_{xx}$ increases with decreasing $T$, while $R_{yy}$ exhibits the opposite trend. For $T > 70$ mK, the 2DHS becomes nearly isotropic, with $R_{yy}$ slightly larger than $R_{xx}$, and both $R_{xx}$ and $R_{yy}$ gradually decrease

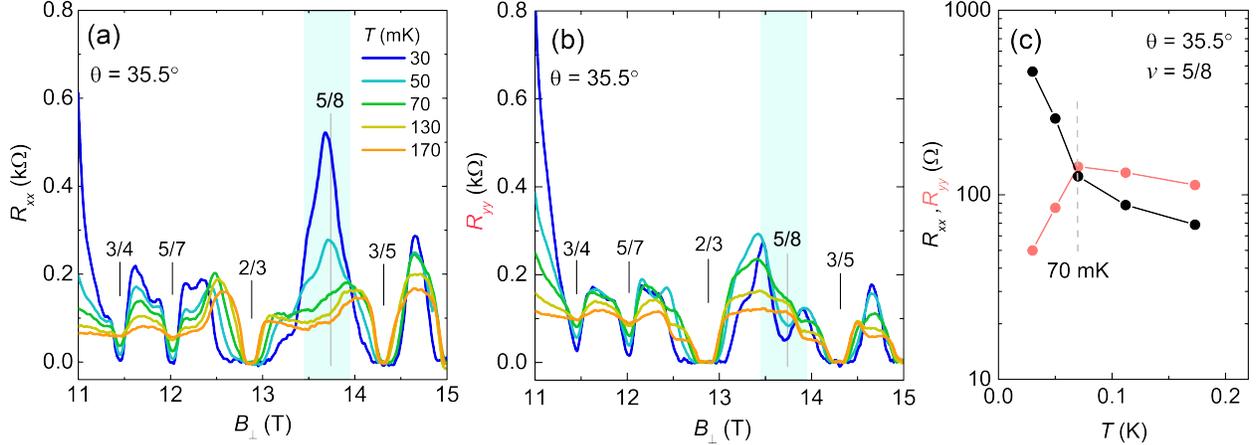

Fig. S2. **Temperature dependence of $R_{xx}$ and $R_{yy}$ in tilted $B$.** (a) $R_{xx}$ and (b) $R_{yy}$ vs $B_\perp$ traces for sample A1 measured at different $T$ in tilted $B$ with $\theta = 35.5°$. At $\nu = 5/8$, $R_{xx}$ exhibits a maximum while $R_{yy}$ shows a minimum. These extrema disappear at $T \geq 70$ mK. (c) $R_{xx}$ (black) and $R_{yy}$ (red) at $\nu = 5/8$ vs $T$.

with increasing $T$. It is worth noting that the $\nu = 5/8$ anisotropic phase in tilted field at $\theta = 35.5°$ disappears above a lower $T \simeq 70$ mK compared to what we observe at $\theta = 0$, suggesting that the anisotropic phase at $\nu = 5/8$ is more fragile in tilted $B$ (with $B_\parallel$ along $[1\bar{1}0]$ direction) than in purely perpendicular $B$. This is in stark contrast to the anisotropic phases in the $N \geq 2$ LLs of GaAs two-dimensional electron systems (2DESs) which survive up to higher temperatures in tilted $B$ [1].

## IV. ROLE OF IN-PLANE MAGNETIC FIELD ALONG $[1\bar{1}0]$ AND $[110]$ DIRECTIONS

Figure S3 summarizes the role of in-plane magnetic field $B_\parallel$ along two mutually perpendicular crystal directions $[1\bar{1}0]$ ($R_{xx}$ direction) and $[110]$ ($R_{yy}$ direction). Note that data in Figs. S3(a-c) and in Figs. S3(d-f) were measured for two different samples (A1 and A2) from the same wafer, using different contact geometries. Here we only focus on and compare the qualitative trends near $\nu = 5/8$. At $\theta = 0$, both samples exhibit an anisotropic behavior near $\nu = 5/8$ with $R_{xx} \ll R_{yy}$ [Figs. S3(a,d)]. At $\theta = 33.4°$ with $B_\parallel$ along $[1\bar{1}0]$ direction, the hard axis of anisotropy switches to $[1\bar{1}0]$ direction [Fig. S3(b)]. In contrast, at $\theta = 35.2°$ with $B_\parallel$ along $[110]$, the hard-axis direction remains along $[110]$ and the anisotropy near

4

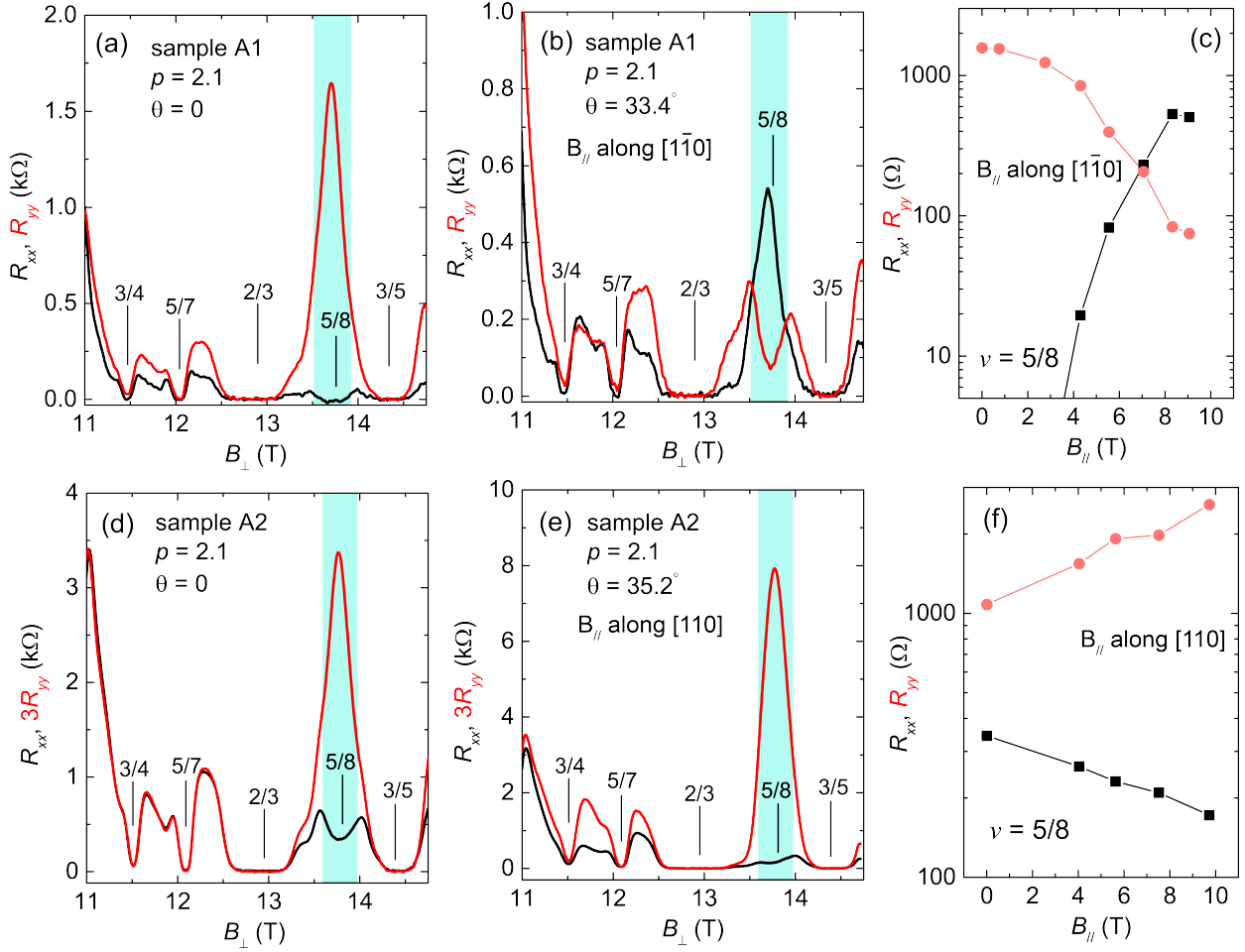

Fig. S3. **Role of in-plane magnetic field.** (a,b) $R_{xx}$ and $R_{yy}$ vs $B_\perp$ traces for sample A1 measured at $\theta = 0°$ and $\theta = 33.4°$. (c) $R_{xx}$ and $R_{yy}$ vs $B_\parallel$. For data in (b,c), $B_\parallel$ is applied along $[1\bar{1}0]$ ($R_{xx}$) direction. (d,e) $R_{xx}$ and $R_{yy}$ vs $B_\perp$ traces for sample A2 measured at $\theta = 0°$ and $\theta = 35.2°$. (f) $R_{xx}$ and $R_{yy}$ vs $B_\parallel$. For data in (e,f), $B_\parallel$ is applied along $[110]$ ($R_{xx}$) direction. Note that in (d,e), the amplitude of $R_{yy}$ traces is multiplied by a factor of three to take into account the effect of contact configuration geometry.

$\nu = 5/8$ becomes larger.

The contrasting behaviors of $R_{xx}$ and $R_{yy}$ vs $B_\parallel$ along $[1\bar{1}0]$ and $[110]$ directions are summarized in Figs. S3(c,f). A large $B_\parallel$ along $[1\bar{1}0]$ can change the hard-axis direction from $[110]$ to $[1\bar{1}0]$, while a large $B_\parallel$ along $[110]$ does not change the hard-axis direction. Instead, the anisotropy at $\nu = 5/8$ gradually grows with increasing $B_\parallel$ along $[110]$. It seems that in a purely perpendicular $B$, there is some native symmetry-breaking field that favors hard axis

along [110]. In GaAs 2DESs, the majority of S/N phases in $N \geq 2$ LLs exhibit a hard axis along the [1$\bar{1}$0] direction, although examples of the hard axis along [110] have been reported under specific conditions [2, 3]. The origin of such internal symmetry breaking field that favors a certain crystal direction over the other remains an unresolved question. Our results show that a large $B_\parallel$ can force the hard axis direction to be parallel to the $B_\parallel$, consistent with Hartree–Fock calculations of the unidirectional charge density wave orientation energy induced by a tilted magnetic field [4].

## V. CALCULATED LANDAU LEVEL FAN DIAGRAM

In GaAs 2DHSs, the larger effective mass ($m^* \simeq 0.5 m_0$, where $m_0$ is the free electron mass) compared to electrons ($m^* \simeq 0.067 m_0$) suppresses the LL separation at high $B_\perp$, leading to a significant mixing of higher LLs into the wavefunction of the lowest LL. This can modify the electron-electron interaction and play an important role in the stabilization of different correlated states in both lowest and excited LLs. Such mixing is quantified by the LL mixing (LLM) parameter $\kappa$, defined as the ratio of Coulomb energy $E_{\text{Coul}}$ and cyclotron energy $E_{cyc}$.

For GaAs 2DESs, LLM is small at high $B_\perp$ ($\kappa \simeq 0.7$ at 14 T). For GaAs 2DHSs, the cyclotron energy is not well defined at high $B_\perp$, making a quantitative estimate of $\kappa$ difficult [5]. Figure S4 shows the energy $E$ vs $B_\perp$ LL fan diagram for our GaAs 2DHS (sample A) calculated in a purely $B_\perp$ using the multiband envelope function approximation and the $8 \times 8$ Kane Hamiltonian [6]. On the right side of Fig. S4, we show the size of half of the Coulomb energy, $e^2/4\pi\epsilon l_B$ where $l_B = (\hbar/eB)^{1/2}$ is the magnetic length, at $\nu = 5/8$. Evidently, the Coulomb energy is significantly larger than the LL separations, underscoring the importance of LLM. In our 2DHSs, the significant LLM modifies the interaction between holes, and hence also modifies the interaction between composite fermions (CFs).

It is worth noting that in addition to the samll LL separations compared to the size of Coulomb energy, the hole LLs also exhibit a complex behavior because of the significant mixing between heavy-hole and light-hole states, and strong spin-orbit coupling. The LLs are highly nonlinear and exhibit numerous crossings. Although our calculated LLs qualitatively capture these key features, a quantitative accuracy is still lacking. For example, similar LL calculations typically predict the LL crossings to occur at a lower $B_\perp$ [7–10]. Therefore, the

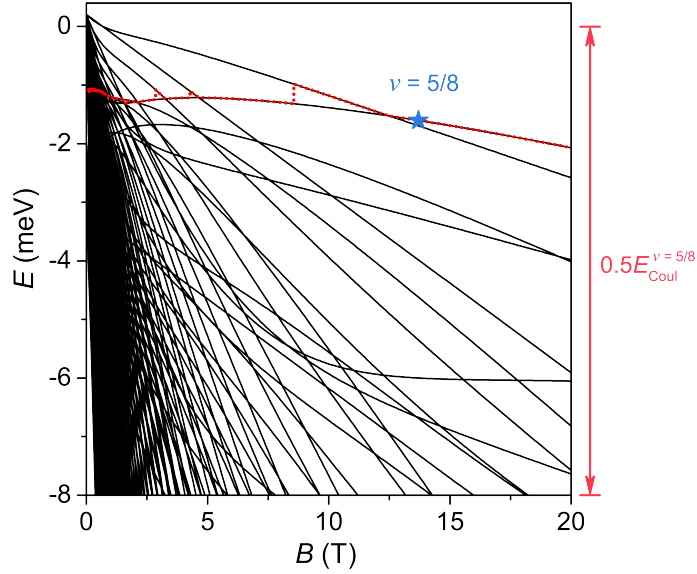

Fig. S4. **Calculated Landau level fan diagram.** Calculated energy $E$ vs $B_\perp$ LL fan diagram for our 2DHS (sample A; $w = 20$ nm and $p = 2.08 \times 10^{11}$ cm$^{-2}$) at $\theta = 0°$. The red dotted line traces the Fermi energy. The position of Fermi energy at $\nu = 5/8$ is marked with a star. On the right, we show the size of half of the Coulomb energy, $e^2/4\pi\epsilon l_B$ where $l_B = (\hbar/eB)^{1/2}$ is the magnetic length, at $\nu = 5/8$.

calculated LLs in Fig. S4 should only be treated qualitatively.

## VI.    DATA FOR SAMPLES B AND C

Figure S5 shows the Hall resistance $R_{xy}$ and longitudinal resistances, $R_{xx}$ and $R_{yy}$, vs $B_\perp$ for samples B and C. In both samples, we observe strong FQHSs at $\nu = 2/3$ and $3/5$, evinced by quantized Hall plateaus and vanishing $R_{xx}$ and $R_{yy}$. Near $\nu = 5/8$, an anisotropic behavior in $R_{xx}$ and $R_{yy}$, qualitatively similar to what we see in samples A1 and A2, is observed, indicating that the S/N phase of CFs we report is a robust, reproducible phenomenon. Near $B_\perp = 13.6$ T, $R_{xx}$ and $R_{yy}$ exhibit minima or inflection points, and $R_{xy}$ merges into the 2/3 plateau. These features are possibly indications of developing reentrant FQHSs emerging from the formation of CF bubbles. The $B_\perp$ position of the developing reentrant FQHS corresponds to $\nu \simeq 0.644$, which can be mapped to CF Lambda level filling factor $\nu_{\mathrm{CF}} \simeq 2.23$.

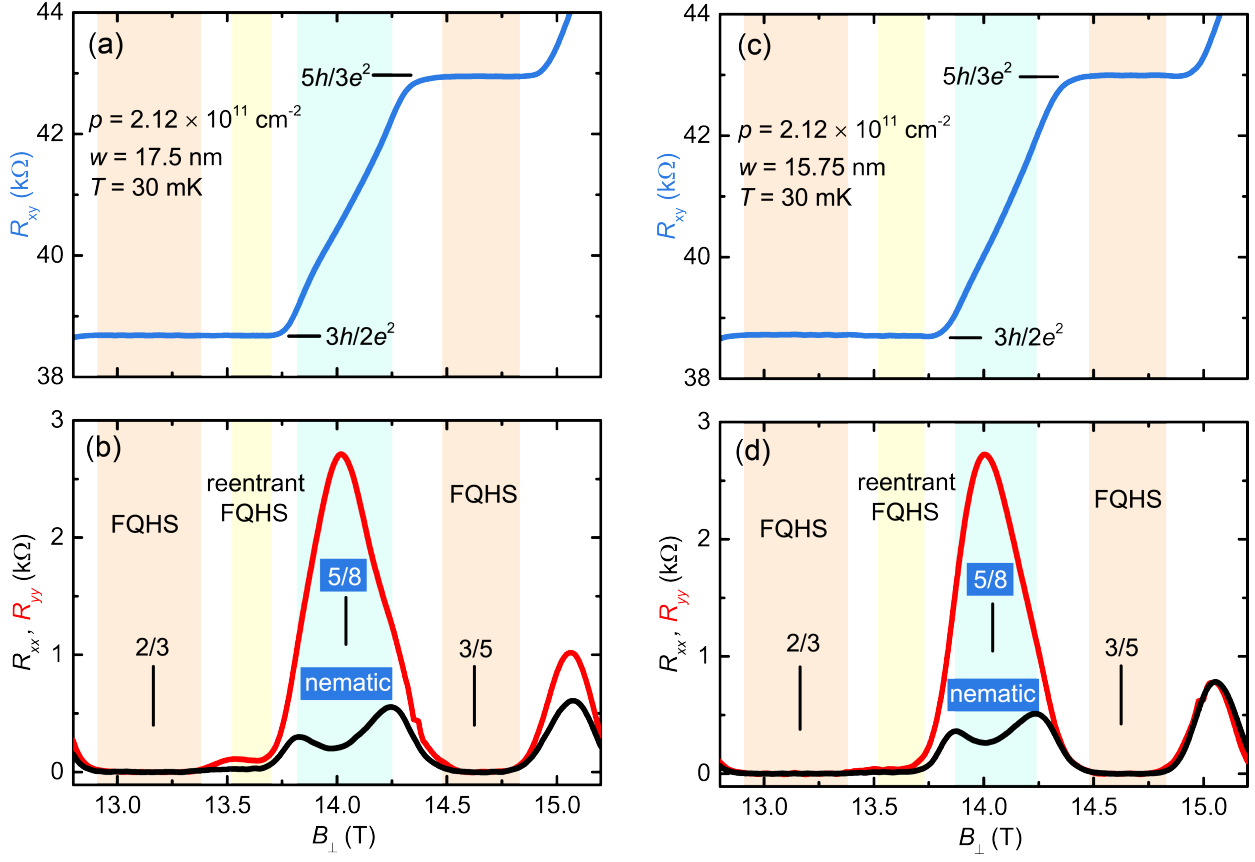

Fig. S5. **Signatures of stripe/nematic phase at $\nu = 5/8$ and reentrant FQHS near $\nu = 2/3$ in samples B and C.** (a) $R_{xy}$ (blue), and (b) $R_{xx}$ (black) and $R_{yy}$ (red) vs $B_\perp$ traces for samples B. This sample has a 2D hole density of $2.12 \times 10^{11}$ cm$^{-2}$, and a QW width of 17.5 nm. (c) $R_{xy}$, and (b) $R_{xx}$ and $R_{yy}$ vs $B_\perp$ traces for samples C. This sample has a 2D hole density of $2.12 \times 10^{11}$ cm$^{-2}$, and a QW width of 15.75 nm.

In Figs. S6(a,b,e,f), we present temperature dependence of the $R_{xx}$ and $R_{yy}$ vs $B_\perp$ traces for samples B and C. The $B_\perp$ positions for FQHSs at $\nu = 3/4, 5/7, 2/3, 3/5$, and $4/7$, as well as for the S/N phase at $\nu = 5/8$, are marked. The black arrows indicate the $B_\perp$ positions of the developing reentrant FQHS near $\nu = 2/3$. In Figs. S6(c,d,g,h), we summarize the temperature dependence of $R_{xx}$ and $R_{yy}$ for the S/N phase at $\nu = 5/8$ and the reentrant FQHS near $\nu = 2/3$. At $\nu = 5/8$, $R_{xx}$ and $R_{yy}$ significantly deviate from each other and exhibit opposite trends at low $T$, indicating the emerging S/N order [Figs. S6(c,g)]. At $B = 13.62$ T, $R_{xx}$ and $R_{yy}$ qualitatively follow the same trend – with decreasing $T$, $R_{xx}$ and $R_{yy}$ first gradually increase, and then decrease sharply when the reentrant FQHS is

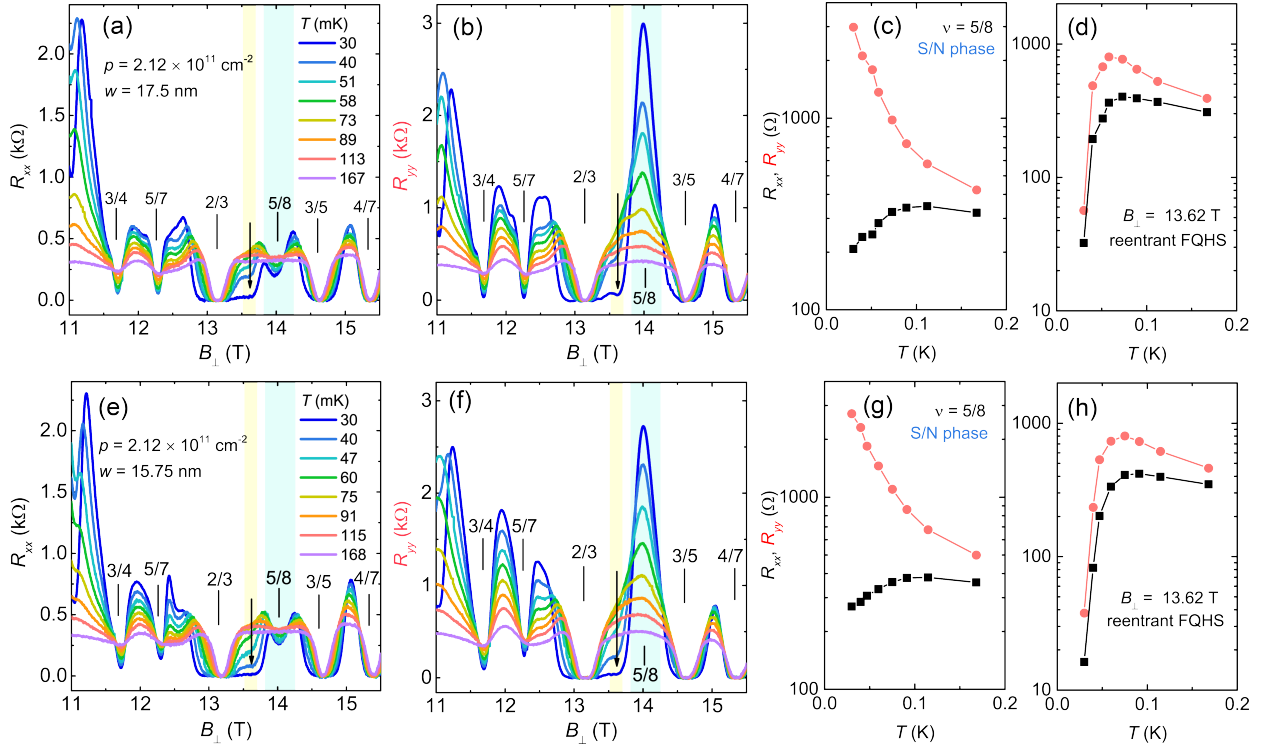

Fig. S6. **Temperature dependence data for samples B and C.** (a, b) $R_{xx}$ and $R_{yy}$ vs $B_\perp$ traces for sample B measured at different $T$. (c, d) $R_{xx}$ and $R_{yy}$ vs $T$: (c) at $\nu = 5/8$ where an S/N phase is observed, and (d) at $B_\perp = 13.62$ T where the 2DHS exhibits a developing reentrant FQHS. (e, f) $R_{xx}$ and $R_{yy}$ vs $B_\perp$ traces for sample C measured at different $T$. (g, h) $R_{xx}$ and $R_{yy}$ vs $T$: (g) at $\nu = 5/8$ where an S/N phase is observed, and (h) at $B_\perp = 13.62$ T where the 2DHS exhibits a developing reentrant FQHS.

developing.

## VII.  DATA FOR SAMPLES D AND E

We also measured two samples with lower hole densities (samples D and E). Figure S7(a) shows data for sample D measured in a purely perpendicular $B$. The 2DHS is nearly isotropic near $\nu = 5/8$, and exhibits a shallow minimum in both $R_{xx}$ and $R_{yy}$. The $B_\perp$ positions of the minima are slightly away from $\nu = 5/8$. At $\theta = 40°$, the 2DHS becomes anisotropic near $\nu = 5/8$ [Fig. S7(b)]: $R_{xx}$ exhibits a peak at $\nu = 5/8$, while $R_{yy}$ shows a minimum.

Figure S8 shows data for sample E whose density is even lower ($p = 0.64$). We do not

9

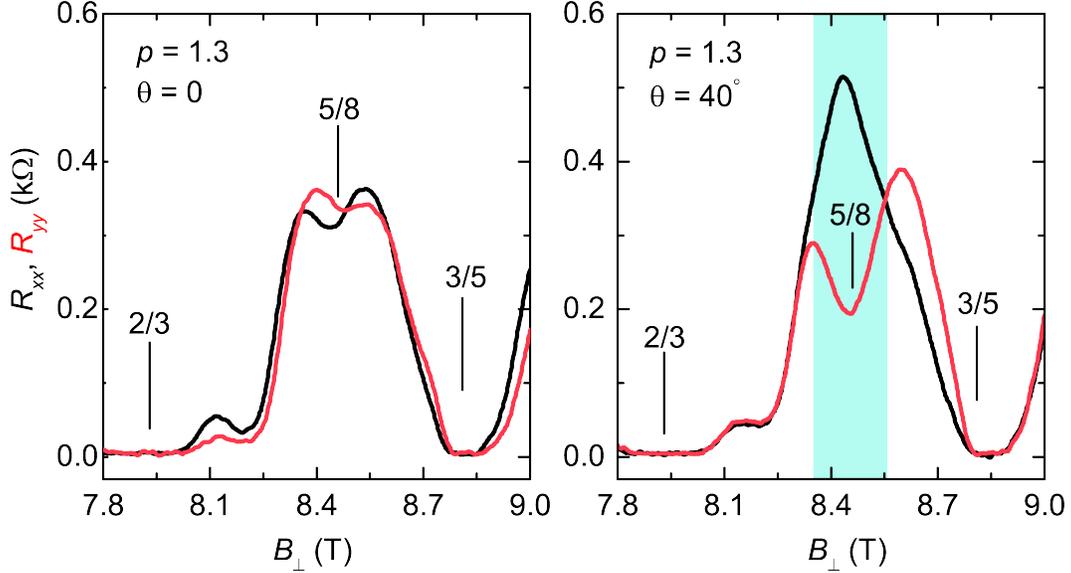

Fig. S7. **Data for sample D without and with tilt.** (a, b) $R_{xx}$ and $R_{yy}$ vs $B_\perp$ traces near $\nu = 5/8$ for sample D, measured at $T = 30$ mK for $\theta = 0$ and $40°$. This sample has a 2D hole density of $1.3 \times 10^{11}$ cm$^{-2}$, and a QW width of 20 nm.

observe significant anisotropic behavior near $\nu = 5/8$. Instead, both $R_{xx}$ and $R_{yy}$ exhibit a minimum centered at $\nu = 5/8$. The minimum may be precursor to a developing FQHS at this filling.

The absence of anisotropy near $\nu = 5/8$ in sample E (and in sample D when $\theta = 0$) is puzzling, because LLM, which we believe is the key to the enhancement of residual CF-CF interaction, is generally more significant at lower densities. However, the highly nonlinear LL fan diagram (e.g., see Fig. S4) and mixed spinor components in the hole LLs add more complexity to the role of 2D hole density in modifying hole-hole interaction and residual CF-CF interaction [10]. Another possible explanation could be that CFs are not fully spin polarized at $\nu = 5/8$ in these lower density samples because $B_\perp$ at $\nu = 5/8$ is relatively lower. In this case, the relevant half-filled CF $\Lambda$L does not have a high orbital index, potentially leading to a different ground state, e.g. an FQHS or a CF Fermi sea. However, one should be cautious because, unlike the simple spin in GaAs 2DESs, strictly speaking, spin is not a good quantum number in GaAs 2DHSs. This is because of the significant heavy-hole light-hole mixing and spin-orbit coupling, which mix hole states with different orbital and spin indices [6]. The spin we refer to here is essentially the LL-pseudospin degree of freedom.

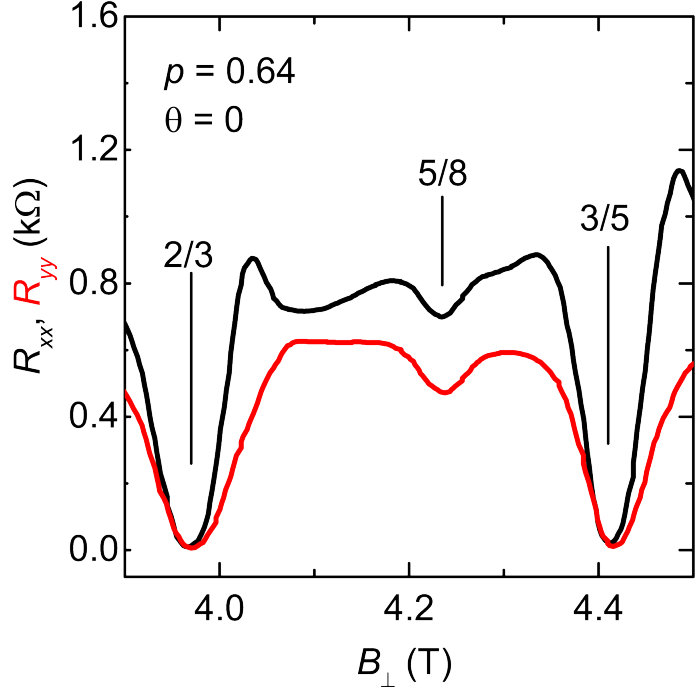

Fig. S8. **Data for sample E.** $R_{xx}$ and $R_{yy}$ vs $B_\perp$ traces near $\nu = 5/8$ for sample E, measured under purely perpendicular $B$. This sample has a 2D hole density of $0.64 \times 10^{11}$ cm$^{-2}$, and a QW width of 20 nm.

Although spin of CFs has been extensively studied in GaAs 2DESs, little is known about the pseudospin polarization of CFs in the FQHS regime of GaAs 2DHSs [11].

It is also worth noting that many-body states in general become more fragile at lower densities because they essentially emerge from Coulomb interaction which becomes weaker at lower $B_\perp$. We emphasize that studying many-body states in low-density samples is challenging, and that lower temperatures and better sample quality are required to further investigate the ground state at $\nu = 5/8$ at low densities.

The emergence of an anisotropic phase at $\nu = 5/8$ at $\theta = 40°$ (sample D) is also interesting. One possibility is that CFs at $\nu = 5/8$ may become fully pseudospin polarized at $\theta = 40°$, leading to an anisotropic phase similar to what we observe in samples A, B and C. Alternatively, it may also be a tilt-induced anisotropic phase of CFs in the $N_{CF} = 1$ ΛL, analogous to the tilt-induced anisotropic phase reported in GaAs 2DESs at $\nu = 5/2$ [12]. The role of in-plane magnetic field in tilted $B$ can be more complex because it can also change the LL fan diagram and spinor composition of the LLs. However, calculations of LLs




\begin{thebibliography}{1}

\bibitem{Tsui.PRL.1982} D. C. Tsui, H. L. Stormer, and A. C. Gossard, Two-Dimensional Magnetotransport in the Extreme Quantum Limit, Phys. Rev. Lett. {\bf 48}, 1559 (1982).

\bibitem{Laughlin.PRL.1983} R. B. Laughlin, Anomalous Quantum Hall Effect: An Incompressible Quantum Fluid with Fractionally Charged Excitations, Phys. Rev. Lett. {\bf 50}, 1395 (1983).

\bibitem{Halperin.Book.2020} \textit{Fractional Quantum Hall Effects: New Developments}, edited by B. I. Halperin and J. K. Jain (World Scientific, Singapore, 2020).

\bibitem{Jain.PRL.1989} J. K. Jain, Composite-fermion approach for the fractional quantum Hall effect, Phys. Rev. Lett. {\bf 63}, 199 (1989).

\bibitem{Halperin.PRB.1993} B. I. Halperin, P. A. Lee, and N. Read, Theory of the half-filled Landau level, Phys. Rev. B {\bf 47}, 7312 (1993).

\bibitem{Jain.Book.2007} J. K. Jain, \textit{Composite fermions}, (Cambridge University Press, Cambridge, England, 2007).
\bibitem{Willett.PRL.1993} R. L. Willett, R. R. Ruel, K. W. West, and L. N. Pfeiffer, Experimental demonstration of a Fermi surface at one-half filling of the lowest Landau level, Phys. Rev. Lett. {\bf 71}, 3846 (1993).

\bibitem{Kang.PRL.1993} W. Kang, H. L. Stormer, L. N. Pfeiffer, K. W. Baldwin, and K. W. West, How real are composite fermions?, Phys. Rev. Lett. {\bf 71}, 3850 (1993).

\bibitem{Kamburov.PRL.2014} D. Kamburov, Yang Liu, M. A. Mueed, M. Shayegan, L. N. Pfeiffer, K. W. West, and K. W. Baldwin, What Determines the Fermi Wave Vector of Composite Fermions?, Phys. Rev. Lett. {\bf 113}, 196801 (2014). 

\bibitem{Willett.PRL.1987} R. Willett, J. P. Eisenstein, H. L. Störmer, D. C. Tsui, A. C. Gossard, and J. H. English, Observation of an even-denominator quantum number in the fractional quantum Hall effect, Phys. Rev. Lett. {\bf 59}, 1776 (1987).

\bibitem{Moore.NPB.1991} G. Moore and N. Read, Nonabelions in the fractional quantum hall effect, Nucl. Phy. B {\bf 360(2-3)}, 362-396 (1991).

\bibitem{Read.PRB.2000} N. Read and Dmitry Green, Paired states of fermions in two dimensions with breaking of parity and time-reversal symmetries and the fractional quantum Hall effect, Phys. Rev. B {\bf 61}, 10267 (2000).

\bibitem{Sharma.PRB.2020} Anirban Sharma, Songyang Pu, and J. K. Jain, Bardeen-Cooper-Schrieffer pairing of composite fermions, Phys. Rev. B 104, 205303 (2020).

\bibitem{Lilly.PRL.1999} M. P. Lilly, K. B. Cooper, J. P. Eisenstein, L. N. Pfeiffer, and K. W. West, Evidence for an Anisotropic State of Two-Dimensional Electrons in High Landau Levels, Phys. Rev. Lett. {\bf 82}, 394 (1999).

\bibitem{Du.SSC.1999} R. R. Du, D. C. Tsui, H. L. Stormer, L. N. Pfeiffer, K. W. Baldwin, and K. W. West, Strongly anisotropic transport in higher two-dimensional Landau levels, Solid State Commun. {\bf 109}, 389 (1999).

\bibitem{Koulakov.PRL.1996} A. A. Koulakov, M. M. Fogler, and B. I. Shklovskii, Charge Density Wave in Two-Dimensional Electron Liquid in Weak Magnetic Field, Phys. Rev. Lett. {\bf 76}, 499 (1996).

\bibitem{Fogler.PRB.1996} M. M. Fogler, A. A. Koulakov, and B. I. Shklovskii, Ground state of a two-dimensional electron liquid in a weak magnetic field, Phys. Rev. B {\bf 54}, 1853 (1996).

\bibitem{Moessner.PRB.1996} R. Moessner and J. T. Chalker, Exact results for interacting electrons in high Landau levels, Phys. Rev. B {\bf 54}, 5006 (1996).

\bibitem{Fradkin.PRB.1999} Eduardo Fradkin, Steven A. Kivelson, Liquid-crystal phases of quantum Hall systems, Phys. Rev. B {\bf 59}, 8065 (1999).

\bibitem{Fradkin.PRL.2000} Eduardo Fradkin, Steven A. Kivelson, Efstratios Manousakis, and Kwangsik Nho, Nematic Phase of the Two-Dimensional Electron Gas in a Magnetic Field, Phys. Rev. Lett. {\bf 84}, 1892 (2000).

\bibitem{Oganesyan.PRB.2001} Vadim Oganesyan, Steven A. Kivelson, and Eduardo Fradkin, Quantum theory of a nematic Fermi fluid, Phys. Rev. B {\bf 64}, 195109 (2001).

\bibitem{Fradkin.review.2010} Eduardo Fradkin, Steven A. Kivelson, Michael J. Lawler, James P. Eisenstein, and Andrew P. Mackenzie, Nematic Fermi Fluids in Condensed Matter Physics,  Annu. Rev. Condens. Matter Phys. {\bf 1}, 153 (2010).

\bibitem{Lee.PRL.2018} Kyungmin Lee, Junping Shao, Eun-Ah Kim, F. D. M. Haldane, and Edward H. Rezayi, Pomeranchuk Instability of Composite Fermi Liquids, Phys. Rev. Lett. {\bf 121}, 147601 (2018).

\bibitem{Sammon.PRB.2019} See also M. Sammon, X. Fu, Yi Huang, M. A. Zudov, B. I. Shklovskii, G. C. Gardner, J. D. Watson,
M. J. Manfra, K. W. Baldwin, L. N. Pfeiffer, and K. W. West, Resistivity anisotropy of quantum Hall stripe phases, Phys. Rev. B {\bf 100} 241303(R) (2019).

\bibitem{Chung.NatMater.2021} Yoon Jang Chung, K. A. Villegas Rosales, K. W. Baldwin, P. T. Madathil, K. W. West, M. Shayegan and L. N. Pfeiffer, Ultra-high-quality two-dimensional electron systems, Nat. Mater. {\bf 20}, 632-637 (2021).

\bibitem{Chung.PRM.2022} Yoon Jang Chung, C. Wang, S. K. Singh, A. Gupta, K. W. Baldwin, K. W. West, R. Winkler, M. Shayegan, and L. N. Pfeiffer, Record-quality GaAs two-dimensional hole systems, Phys. Rev. Mater. {\bf 6}, 034005 (2022).

\bibitem{Gupta.PRM.2024} Adbhut Gupta, C. Wang, S. K. Singh, K. W. Baldwin, R. Winkler, M. Shayegan, and L. N. Pfeiffer, Ultraclean two-dimensional hole systems with mobilities exceeding $10^7$ cm$^2$/Vs, Phys. Rev. Mater. {\bf 8}, 014004 (2024).

\bibitem{Footnote.direction} Here [1$\bar{1}$0] is defined as the crystal direction perpendicular to the major flat of the GaAs wafer.

\bibitem{SM} See Supplemental Material,which includes Refs. \cite{Zhu.PRL.2002, Pollanen.PRB.2015, Jungwirth.PRB.1999, Wang.PRB.2025, Winkler.Book.2003, Liu.PRB.2014, Lupatini.PRL.2020, Ma.PRL.2022, Wang.RPP.2025, Huang.PRL.2019}, for additional magneto-transport data and the calculated Landau level fan diagram for our GaAs 2DHS.

\bibitem{Zhu.PRL.2002} J. Zhu, W. Pan, H. L. Stormer, L. N. Pfeiffer, and K. W. West, Density-Induced Interchange of Anisotropy Axes at Half-Filled High Landau Levels, Phys. Rev. Lett. {\bf 88}, 116803 (2002).

\bibitem{Pollanen.PRB.2015} J. Pollanen, K. B. Cooper, S. Brandsen, J. P. Eisenstein, L. N. Pfeiffer, and K. W. West, Heterostructure symmetry and the orientation of the quantum Hall nematic phases, Phys. Rev. B {\bf 92}, 115410 (2015).

\bibitem{Jungwirth.PRB.1999} T. Jungwirth, A. H. MacDonald, L. Smrčka, S. M. Girvin, Field-tilt anisotropy energy in quantum Hall stripe states, Phys. Rev. B {\bf 60}, 15574 (1999).

\bibitem{Wang.PRB.2025} Chengyu Wang, A. Gupta, S. K. Singh, L. N. Pfeiffer, K. W. Baldwin, R. Winkler, and M. Shayegan, Competing many-body phases at small fillings in ultrahigh-quality GaAs two-dimensional hole systems: Role of Landau level mixing, Phys. Rev. B {\bf 111}, 085429 (2025).

\bibitem{Winkler.Book.2003} R. Winkler,
\textit{Spin-Orbit Coupling Effects in Two-Dimensional Electron and Hole Systems}, (Springer, Berlin, 2003).

\bibitem{Liu.PRB.2014} Yang Liu, S. Hasdemir, D. Kamburov, A. L. Graninger, M. Shayegan, L. N. Pfeiffer, K. W. West, K. W. Baldwin, and R. Winkler, Even-denominator fractional quantum Hall effect at a Landau level crossing, Phys.Rev. B {\bf 89}, 165313 (2014).

\bibitem{Lupatini.PRL.2020} M. Lupatini, P. Knüppel, S. Faelt, R. Winkler, M. Shayegan, A. Imamoglu, and W. Wegscheider, Spin Reversal of a Quantum Hall Ferromagnet at a Landau Level Crossing, Phys. Rev. Lett. {\bf 125}, 067404 (2020).

\bibitem{Ma.PRL.2022} Meng K. Ma, Chengyu Wang, Y. J. Chung, L. N. Pfeiffer, K. W. West, K. W. Baldwin, R. Winkler, and M. Shayegan, Robust Quantum Hall Ferromagnetism near a Gate-Tuned $\nu=1$ Landau Level Crossing, Phys. Rev. Lett. {\bf 129}, 196801 (2022).

\bibitem{Wang.RPP.2025} Chengyu Wang, A. Gupta, S. K. Singh, C. T. Tai, L. N. Pfeiffer, K. W. Baldwin, R. Winkler, and M. Shayegan, Even-denominator fractional quantum Hall states with spontaneously broken rotational symmetry, Rep. Prog. Phys. {\bf 88} 100501 (2025).

\bibitem{Huang.PRL.2019} Ke Huang, P. Wang, L. N. Pfeiffer, K. W. West, K. W. Baldwin, Y. Liu, X. Lin, Resymmetrizing Broken Symmetry with Hydraulic Pressure, Phys. Rev. Lett. {\bf 123}, 206602 (2019).

\bibitem{Wang.PRL.2022} Chengyu Wang, A. Gupta, S. K. Singh, Y. J. Chung, L.N. Pfeiffer, K. W. West, K. W. Baldwin, R. Winkler, and M. Shayegan, Even-Denominator Fractional Quantum Hall State at Filling Factor $\nu=3/4$, Phys. Rev. Lett. {\bf 129}, 156801 (2022).

\bibitem{Wang.PNAS.2023} Chengyu Wang, Adbhut Gupta, Pranav T. Madathil, Siddharth K. Singh, Yoon Jang Chung, Loren N. Pfeiffer loren, Kirk W. Baldwin, and Mansour Shayegan, Next-generation even-denominator fractional quantum Hall states of interacting composite fermions, Proc. Nat. Acad. Sci. {\bf 120}, e2314212120 (2023). 

\bibitem{Lee.PRB.2002} Seung-Yeop Lee, Vito W. Scarola, and J.K. Jain, Structures for interacting composite fermions: Stripes, bubbles, and fractional quantum Hall effect, Phys. Rev. B {\bf 66}, 085336 (2002)

\bibitem{Goerbig.PRL.2004} M. O. Goerbig, P. Lederer, and C. Morais Smith, Possible Reentrance of the Fractional Quantum Hall Effect in the Lowest Landau Level, Phys. Rev. Lett. {\bf 93}, 216802 (2004).

\bibitem{Shingla.NatPhys.2023} Vidhi Shingla, Haoyun Huang, Ashwani Kumar, Loren N. Pfeiffer, Kenneth W. West, Kirk W. Baldwin, and Gábor A. Csáthy, A highly correlated topological bubble phase of composite fermions, Nat. Phys. {\bf 19}, 689 (2023).

\bibitem{Ro.PRB.2020} Dohyung Ro, S. A. Myers, N. Deng, J. D. Watson, M. J. Manfra, L. N. Pfeiffer, K. W. West, and G. A. Csáthy, Stability of multielectron bubbles in high Landau levels, Phys. Rev. B {\bf 102}, 115303 (2020).

\bibitem{Footnote.config} Note that the circuit configurations we use here [insets in Figs. \ref{fig:Tdep}(a,b)] are different from those in Fig. \ref{fig:fullfield}, but the traces are qualitatively similar. Because of the geometric effects, in Figs. \ref{fig:Tdep}(a,b), the $R_{xx}$ and $R_{yy}$ values are overall smaller and the anisotropy at $\nu=5/8$ is larger compared to data in Fig. \ref{fig:fullfield} \cite{Simon.PRL.1999}.

\bibitem{Simon.PRL.1999} Steven H. Simon, Comment on “Evidence for an Anisotropic State of Two-Dimensional Electrons in High Landau Levels”, Phys. Rev. Lett. {\bf 83}, 4223 (1999).

\bibitem{Cooper.PRB.2002} K. B. Cooper, M. P. Lilly, J. P. Eisenstein, L. N. Pfeiffer, and K. W. West, Onset of anisotropic transport of two-dimensional electrons in high Landau levels: Possible isotropic-to-nematic liquid-crystal phase transition, Phys. Rev. B {\bf 65}, 241313(R) (2002).

\bibitem{Footnote.7/12} While the anisotropy at $\nu=7/12$ increases with decreasing $T$, no qualitative difference is seen at this filling in $R_{xx}$ and $R_{yy}$ traces --- both show a peak in the whole range of $T$ [Figs. \ref{fig:Tdep}(a,b)].

\bibitem{Madathil.PRL.2023} P. T. Madathil, K. A. Villegas Rosales, C. T. Tai, Y. J. Chung, L. N. Pfeiffer, K. W. West, K. W. Baldwin, and M. Shayegan, Delocalization and Universality of the Fractional Quantum Hall Plateau-to-Plateau Transitions, Phys. Rev. Lett. {\bf 130}, 226503 (2023).

\bibitem{Lilly.PRL.1999.tilt} M. P. Lilly, K. B. Cooper, J. P. Eisenstein, L. N. Pfeiffer, and K. W. West, Anisotropic States of Two-Dimensional Electron Systems in High Landau Levels: Effect of an In-Plane Magnetic Field, Phys. Rev. Lett. {\bf 83}, 824 (1999).

\bibitem{Lee.PRL.2001} Seung-Yeop Lee, Vito W. Scarola, and J. K. Jain, Stripe Formation in the Fractional Quantum Hall Regime, Phys. Rev. Lett. {\bf 87}, 256803 (2001).

\bibitem{Scarola.PRL.2002} V. W. Scarola, J. K. Jain, and E. H. Rezayi, Possible Pairing-Induced Even-Denominator Fractional Quantum Hall Effect in the Lowest Landau Level, Phys. Rev. Lett. {\bf 88}, 216804 (2002). 

\bibitem{Goldman.SurfSci.1990} V. J. Goldman and M. Shayegan, Fractional quantum Hall states at $\nu=$ 7/11 and 9/13, Surf. Sci. {\bf 229}, 10 (1990).

\bibitem{Pan.PRL.2003} W. Pan, H. L. Stormer, D. C. Tsui, L. N. Pfeiffer, K. W. Baldwin, and K. W. West, Fractional Quantum Hall Effect of Composite Fermions, Phys. Rev. Lett. {\bf 90}, 016801 (2003).

\bibitem{Liu.PRL.2014} Yang Liu, D. Kamburov, S. Hasdemir, M. Shayegan, L. N. Pfeiffer, K. W. West, and K. W. Baldwin, Phys. Rev. Lett. {\bf 113}, 246803 (2014).

\bibitem{Kumar.NC.2018} Manohar Kumar, Antti Laitinen, and Pertti Hakonen, Unconventional fractional quantum Hall states and Wigner crystallization in suspended Corbino graphene, Nat. Commun. {\bf 9}, 2776 (2018).

\bibitem{Manfra.PRL.2007} M. J. Manfra, R. de Picciotto, Z. Jiang, S. H. Simon, L. N. Pfeiffer, K. W. West, and A. M. Sergent, Impact of Spin-Orbit Coupling on Quantum Hall Nematic Phases, Phys. Rev. Lett. {\bf 98}, 206804 (2007).

\bibitem{Wang.PRL.2023} Chengyu Wang, A. Gupta, S. K. Singh, P. T. Madathil, Y. J. Chung, L. N. Pfeiffer, K. W. Baldwin, R. Winkler, and M. Shayegan, Fractional Quantum Hall State at Filling Factor $\nu=1/4$ in Ultra-High-Quality GaAs Two-Dimensional Hole Systems, Phys. Rev. Lett. {\bf 131}, 266502 (2023).

\bibitem{Zhao.PRL.2023} Tongzhou Zhao, Ajit C. Balram, and J. K. Jain, Composite Fermion Pairing Induced by Landau Level Mixing, Phys. Rev. Lett. {\bf 130}, 186302 (2023).

\bibitem{data} C. Wang, S. K. Singh, C. T. Tai, A. Gupta, L. N. Pfeiffer, K. W. Baldwin, and M. Shayegan, Supporting Data for: "Observation of a Stripe/nematic Phase of Composite Fermions", 10.5281/zenodo.17868425 (2025).

\end{thebibliography}
\end{document}